# WHAT WE SHOULD TEACH, BUT DON'T: PROPOSAL FOR A CROSS POLLINATED HCI-SE CURRICULUM

*Pardha S. Pyla, Manuel A. Pérez-Quiñones, James D. Arthur and H. Rex Hartson*[1]

***Abstract** – Software engineering (SE) and usability engineering (UE), as disciplines, have reached substantial levels of maturity. Each of these two disciplines is now well represented with respect to most computer science (CS) curricula. But, the two disciplines are practiced almost independently – missing opportunities to collaborate, coordinate and communicate about the overall design – and thereby contributing to system failures. Today, a confluence of several ingredients contribute to these failures: the increasing importance of the user interface (UI) component in the overall system, the independent maturation of the human computer interaction area, and the lack of a cohesive process model to integrate the UI experts' usability engineering (UE) development efforts with that of software engineering (SE). This in turn, we believe, is a result of a void in computing curricula: a lack of education and training regarding the importance of communication, collaboration and coordination between the SE and UE processes. In this paper we describe the current approach to teaching SE and UE and its shortcomings. We identify and analyze the barriers and issues involved in developing systems having substantial interactive components. We then propose four major themes of learning for a comprehensive computing curriculum integrating SE, UE, and system architectures in a project environment.*

## INTRODUCTION

For almost half a century, software engineering (SE) has been researched and practiced in various domains and different scales. Even though much younger, usability engineering (UE) has also seen substantial amounts of research and application in a wide variety of user interfaces (UIs). In spite of the commendable level of maturity achieved by these two disciplines independently, software systems with interactive components still fall far short of being perfect. We believe that this is because of the lack of an integrated approach toward developing interactive systems. One of the most important reasons why SE as a field has failed to connect to the UE discipline is that the processes and activities in SE have their roots in non-interactive, batch processing based, systems-side applications development. On the other hand, one plausible explanation for why the UE discipline failed to connect to the SE domain is a lack of knowledge on the part of the usability engineers toward the SE processes, constraints and schedules. Overall, the developers from these two disciplines have very little understanding and appreciation for each other's skills, techniques and backgrounds.

In recent times, the addition of complex user interfaces has changed the nature of software systems. Now more than ever, the average end-users' expectations from software systems has increased significantly. Moreover, the massive proliferation of affordable computing resources to the general public has increased the number of 'novice' computer users to an all time high. Today, from the user's perspective, the interface *is* the system [1]. In effect, user interfaces have the power to "make or break" a software product [2]. The importance of user interfaces in current day systems, and the lack of theoretical foundations in software engineering to support the development of this component have resulted in a dangerous situation: the lack of good usability in software products.

Anecdotal evidence suggests that a major number of software systems that appear to be free of software errors or bugs still have significant usability problems. This evidence, together with the fact that user interface code is often 50% or more [3] of a project, presents an alarming picture in the development of interactive software systems. The fundamental reason for this state-of-the-art is the lack of an integrated process model to provide a development infrastructure in which the UE and SE life cycles co-exist in complementary roles [4]. We argue that this is due to the absence of computer science curricula that attempt to teach students of the *existence* and the barriers involved in *coordinating* these two processes. In this paper, we provide the background, motivation, issues, and themes for a cross pollinated computer science curriculum that addresses HCI and SE topics that will help in bridging the gap between the two domains. We then provide a blueprint for a semester long project oriented course that uses these themes.

## BACKGROUND AND MOTIVATION

Interactive software systems have both functional (non-user interface) and user interface parts. Although the separation of code into two clearly identifiable modules is not always possible, the two parts exist conceptually and each must be designed on its own merits.

[1] Pardha S. Pyla, Manuel A. Pérez-Quiñones, James D. Arthur and H. Rex Hartson, Department of Computer Science, Virginia Tech, 660 McBryde Hall, Blacksburg, VA 24061, {ppyla, perez, arthur, hartson}@cs.vt.edu

Frequently, software development efforts leave the user interface design for the end of the project. What is not understood is that requirements for the user interface have deep consequences on the design of the functional part of the system and even on the system architecture [5]. Moreover, each development group gathers its own requirements, resulting in separate requirements for the user interface and the functional core. But in reality, the two groups should work *together* to gather a single set of requirements - the requirements of the system. Those should then be viewed by the developer with regard to the implications for each software component.

Also, the implementation of most graphical user interfaces have an interesting characteristic: the control thread of the program is taken out of the functional core and placed in system libraries. The functional core no longer controls what function is called next. These control decisions are now made based on the results of user initiated actions on the interface. Therefore, the application becomes an event-based system. Such a seemingly simple characteristic places many restrictions on the software functionality.

Regretfully, the situation gets worse each day. The World Wide Web has popularized client-server architectures. Now software developers not only have to contend with event-based threads of control, but also with client-server considerations. This often requires the use of multiple languages (e.g. HTML, CSS, JavaScript, PHP, SQL) to implement the client and server components in a single project. For example, a simple form on a web page that requires validation can involve three languages: HTML, JavaScript and a server language. This makes traditional software engineering concerns such as coupling, cohesion, and modularity very challenging in their own right. To achieve the goals for both parts of an interactive system, i.e., to create an efficient and reliable system with required functionality and high usability, effective development processes are required for both the UE and SE lifecycles. They must include well drawn communication, collaboration, and coordination lines that ultimately help in synchronizing, identifying and realizing constraints and dependencies between the two development processes [4].

We believe such cooperative existence of the two domains is proving difficult because of the lack of an education curriculum in computer science and/or software engineering to prepare tomorrow's software professionals with the knowledge of the existence of the two lifecycles, the issues involved in coordinating them, and the necessary training to work in such a collaborative setting.

## CURRENT ACADEMIC ENVIRONMENT AND RELATED WORK

We recognize that there are efforts being made to bring the user interface and functional development methodologies together, but we argue that we are not yet there. This is because the development of interactive systems should not be done with a bias towards either the user interface or the functional parts of the system, but should be performed with equal representation for the two.

Traditionally, the concepts, theory and techniques associated with the user interface (UI) design domain are taught as a part of Human Computer Interaction (HCI) courses in most computer science departments. These courses use text books such as [6-8] and cover topics that are completely focused on the user interface side of system development. Little or no information is provided on the software architectural implications that arise from user interface design decisions. For example, Rosson and Carroll talk about scenarios and focus on the UE life cycle and process [7]. They do not provide a clear description of the two development efforts and the implications for the functional core. The incremental case study provided in this book proceeds through different interface design stages (as the chapters progress), and ends up with a system entirely developed and including user documentation. The readers are not informed how the user interface specifications are communicated to the software developers, nor how the design constraints are negotiated between the two development bodies.

Students who complete this type of academic HCI course gain little, if any appreciation for the development process associated with the functional part of the system. They may wrongly believe that any usability or user interface specification derived will 'somehow' get automatically implemented by 'someone', which in reality is rarely the case. To the contrary, simple usability features, such as undo, that the students might take for granted, can have serious architectural implications in the functional part of the system [5].

The SE courses as they are offered in most computer science departments are no better than their HCI counterparts when it comes to addressing the connections between the two domains. SE courses often omit any references to user interaction development techniques. We surveyed nine out of 13 of the top selling books in software engineering/software design and engineering (according to facultyonline.com website) [9]. Out of the nine, only three books identify and dedicate sections to user interface development. Even so, those three still do not address the dependencies and implications of the two processes on one another. Students completing software engineering courses that follow the syllabus in textbooks like these obviously will gain little or no appreciation for the user interface issues and the constraints on the functional core that arise due to them.

Some textbooks have attempted to address this issue, but their impact in changing the academic environment has been minimal. For example, Hix and Hartson [1] discuss at length the connections that should to exist between UE and the rest of the SE lifecycle. However, because their book focuses on usability engineering methods and techniques and does not suggest how to integrate the two domains, the SE community is not really aware of its impact.

In recent years there has been a growing awareness of the importance of bridging the gap between the SE and UE domains [4], [10]. There have also been appeals for curricula that integrate these two disciplines [2], [10]. For example, Sefah points out that there are very few software engineers who understand the human-centered design process. He states that one of the reasons for this is the lack of a proper educational framework. He also provides a list of skills that one should have to perform human centered design [11]. Latzina and Rummel reason that because of the lack of HCI studies available for computer science students, much of the usability training is left to corporate training workshops. Because of cost and time factors, these workshops are extremely short (about 2 days) in duration, thereby reducing the course content to "commonplace statements" [12]. Wahl proposes to teach SE students "usability testing" so that "students can learn about user-centered design and what makes software usable by running usability tests" [13]. We believe this focus on usability testing alone to be unsuitably limited as it promotes the idea that UE is just usability testing that is performed at the end of software development, and that this is enough to teach user-centered design.

Leventhal and Barnes have been advocating and implementing a curriculum that integrates HCI and SE within a computer science course that "emphasize(s) some SE notions in the context of HCI concepts" [14], [15]. They incorporate some of the SE topics into a project oriented HCI course. On the other hand, Veer and Vliet appeal for a 'minimal' HCI course to be incorporated into a software curriculum [16] to train students for a more integrated approach towards development of interactive systems. Similarly, the joint task force on computing curricula commissioned by ACM Education Board, IEEE-Computer Society Educational Activities Board, and other professional societies, also recommend bits and pieces of HCI in their SE courses [17]. They recommend a separate HCI course (SE212), similar to a number of pure HCI courses taught in universities, that covers the usability engineering processes, methodologies, architectures and techniques. Unfortunately, their recommendations for core software engineering courses such as "Software process and management" (SE324), "Software Project Management" (SE323), "Software design and architecture" (SE311), etc. do not even mention user interface issues.

Even though Leventhal and Barnes' suggestions are the closest to our own arguments, their approach is more focused on to the HCI component. We propose a more balanced approach toward teaching SE and UE, and not try to incorporate pieces of one lifecycle's methodologies into the other. Moreover, Leventhal and Barnes do not address the issue of dependencies and constraints between the two lifecycles. We consider these to be one of the most important aspects the developers of tomorrow should comprehend.

## BARRIERS

In order to effectively connect the two disciplines of SE and UE, there is an urgent need for a process model that provides a development infrastructure [4] combined with a cross-pollinated curriculum in the computer science education. The need for an integrated process model is paramount because the UE and SE lifecycles should co-exist in complementary roles that produce non-conflicting requirements. Therefore, the need arises for a cross-pollinated curriculum that better prepares the students for the realities of complex processes for developing interactive systems. In this section we describe the issues involved in such a development effort and enumerate the changes that we believe are necessary.

### Barriers to an Integrated Process Model

The adoption of an integrated process model, as referred to in this paper, is fraught with many hurdles. The objectives of SE and UE are achieved by the two developer roles using different activities, development processes, timelines, iterativeness, techniques and focus. Some of the salient hurdles are described below.

- **Differences in requirements representation:** Most requirement specifications documented by software engineers use plain English and are generally very detailed. These specifications are written specifically to drive the design process. On the other hand, usability engineers specify user interface issues such as feedback, screen layout, colors, etc. using artifacts such as prototypes, usage scenarios, and screen sketches. These artifacts are not detailed or complete enough to derive software specifications; instead they require additional refinement and design formulation before implementation. Therefore, they cannot be used directly to drive the software development process.
- **Differences in testing:** Testing in UE takes place frequently at the end of every stage and on a small scale (in the context of a small number of user tasks) before much, if any, software is committed to the user interface. This results in changing the design specifications for the interface throughout the development effort, and thereby constraining or adversely impacting the development of the functional core. The testing in the SE lifecycle is often an independent stage and is primarily done at the very end. An exception to this is the recent emphasis on lightweight models, in particular, extreme programming, which places more emphasis on iterative testing.
- **Differences in terminology:** Even though certain terms in both lifecycles sound similar they often mean different things. For example a (use case) scenario in SE is used to "identify a thread of usage for the system to be constructed (and) provide a description of how the system will be used" [18]. Whereas in UE, a scenario is

"a narrative or story that describes the activities of one or more persons, including information about goals, expectations, actions, and reactions (of persons)" [7].

- **Dependencies and constraints:** Many system requirements have both a user interface and a functional part. When the usability and software engineers gather requirements separately and without communication, it is easy to capture the requirements that are conflicting and incompatible. Software engineers perform functional analysis from the requirements whereas usability engineers perform a hierarchical task analysis. Each task in the task analysis implies the need for corresponding functions in the SE specifications, and each function in the software design reflect the need for support in one or more user tasks in the user interface. Similarly, there is a direct mapping between the usage scenarios on the UE side and the use cases on the SE side. Each of these pairs of products essentially represents the same system in different ways.

# PROPOSED THEMES FOR AN UNDERGRADUATE CURRICULUM

Clearly we cannot continue to offer SE courses where the topics covered only mention UE methods, i.e., where the techniques taught do not include usability evaluation techniques, where the students are unaware of the differences mentioned in the previous section, and where the simulated applications built or designed as part of course projects have only a minimal user interface. We opine that there should be a broad (possibly multi-semester) CS curriculum integrating SE, HCI, system architectures, etc. in a project environment to prepare the students for the realities of real world software development. In answer to the question of how and what this new curriculum should offer, we propose the following four major themes:

- **Process models and components:** This category covers the topics related to the different process model lifecycles in SE and UE, with the individual stages in each, the timelines, interdependence, and the documents or artifacts that are generated at the end of each stage. This theme should provide an understanding of the individual as well as the integrated SE and UE development process.
- **Techniques and tools:** Techniques are different types of procedures (e.g. heuristic evaluation, regression testing, etc.) that can be used for each of the lifecycle stages. Tools are the implementation of the "techniques" group (e.g. UML diagrams, programming environments, user interaction screen capture systems, etc.). Students should be introduced to an *integrated* set of tools and techniques that can support SE and UE process activities. Such techniques include the identification of synchronization points during system development where the two development roles should compare their products to see that the two representations (of the system) are consistent. Students should have reasonable exposure to such concepts to be productive and to promote change on the current state of practice in industry.
- **Lessons learned:** This category covers the differences, similarities, dependencies and constraints, between process models, components, techniques, artifacts, timelines, professional backgrounds of developer roles, and skill-sets required for each lifecycle expert. Apart from these, this theme also covers the system development guidelines for both processes. Students should also be introduced to the dependencies between the SE and UE lifecycles using results from recent research efforts, such as [5].
- **Application:** This category covers simulated exercises for developing an application that illustrates the dependencies and constraints between the SE and UE processes by requiring the students to go through the whole development process. One way to implement this theme is to have a semester long project for developing interactive system.

TABLE I:
PROPOSED THEMES TO OVERCOME THE MAJOR BARRIERS FOR INTEGRATED DEVELOPMENT PROCESS

| Themes and hurdles overcome | Differences in requirements representation | Differences in testing | Differences in terminology | Dependencies & constraints |
|---|---|---|---|---|
| Lifecycles & components | X | | | |
| Techniques & tools | | X | | |
| Lessons learned | | X | X | X |
| Application of skills | X | X | X | X |

As shown in Table I, the 'process models and components' theme teaches students the different types of artifacts in each stage and therefore prepares them for the differences in requirements representation hurdle. Techniques and tools help the student in tackling the differences in testing hurdle. The 'lessons learned' theme prepares the student to overcome the differences in testing and terminology, and emphasizes the dependencies and constraints between the SE and UE lifecycles. A simulated application development effort would introduce and train the student in overcoming all the barriers described. Different undergraduate programs can place different emphasis on these themes. For example there can be a course per theme, several courses cross cutting the themes, or integration of these themes into existing courses. However, the key is that the themes should cover both SE and UE.

## Blueprint for an Undergraduate Course

Here we describe a possible instantiation of the four themes described above for a semester long introductory course on developing interactive systems. This course can be taught at a junior level and can be followed up with more advanced ones that focus on particular aspects of the development effort (e.g. a course on requirements engineering alone). The proposed calendar for this course is shown in Table II. The first column charts the week number and the focus of lectures in that week (to show which discipline professor should lecture). The second column briefly lists the topics to be covered in a particular week. The third column shows the progress of project phases throughout the semester with the time allocated for each phase. The last column describes the list of project deliverables.

The first four weeks of the course cover the two lifecycle processes and development stages. The requirements phases of these two lifecycles are covered in enough detail to be able to assign the first phase of the project early on. Once the individual lifecycles are covered, one week is spent on motivation for an integrated system development approach - topics such as communication, synchronization, constraints and dependencies are addressed. Here, the current practices of developing the interface and backend components of the systems independently are discussed, and potential problems associated with this approach are explained (e.g. inconsistent requirements, system architectures, non-discovery of such inconsistencies till later in the development life cycle and the high costs associated to fix them). After the motivation for the integration phase, the students should have gained an appreciation for the need for integration, and continue the rest of the project phases by communicating and synchronizing the two development cycles. The next six weeks are used to teach the development methods and techniques.

Another important component of this course is the semester long group project. Project exercises are used to make the students apply the concepts taught in class and to document the common issues and lessons learned in such an integrated development setting. In the first class of the semester the students are divided into teams of six (ideally). Each group is further divided into two subgroups: UI subgroup and SE subgroup.

The first phase of the project involves performing the requirements engineering for the selected application. This phase should be performed individually by the UI and SE subgroups of each project group. During the fifth and sixth weeks, when the lecture covers the integration issues, the second phase of the project is assigned: to check the requirements generated by each of the subgroups for consistency. This phase gives the students first hand experience to observe how many differences exist in their two requirements generation efforts for the same project. From then on, the project phases proceed with the regular development activities of each lifecycle with consistency and synchronization activities being emphasized as needed. The students should maintain a list of lessons learned throughout the project development. The last two weeks are used for team project presentations and discussions about the problems encountered, constraints and dependencies identified, and how they were resolved. The final report due at the end of the semester should document all this and should be in the form of a case study.

### Non-Curricular Issues

- **Collaboration among professors:** In research oriented schools this course can be team taught by professors

TABLE II:
PROPOSED COURSE CALENDAR

| Week # (Focus) | Topics covered and involved themes | Project phase and duration | Project deliverables (Application) |
|---|---|---|---|
| 1 (SE) | SE lifecycle **process models and development phases** (enough detail to be able to assign phase 1 of project) | | Divide class into teams for group project and each group into SE and UE sub-groups |
| 2 (SE) | | Phase 0: 1 week | Identify or assign an interactive system to develop as a semester long project |
| 3 (UE) | UE lifecycle **process models and development phases** (enough detail to be able to assign phase 1) | Phase 1: 2 weeks | Requirements for the system by each group (To be performed separately by the SE and UE subgroups) |
| 4 (UE) | | | |
| 5 (Both) | Motivation for integrated approach. **Techniques and tools (T&T)** for integration: synchronization points, examples of constraints and dependencies. | Phase 2: 2 weeks | RE checks between the two subgroups for consistency. List of lessons learned and issues identified. |
| 6 (Both) | | | |
| 7 (UE) | **T & T** for UE: Usage scenarios, HTA, etc. Similarly for SE: Use cases, functional decomposition, etc. | Phase 3: 2 weeks | HTA and scenarios for UI and use cases and functional decomposition for SE |
| 8 (SE) | | | |
| 9 (UE) | **T & T** for UI conceptual design, screen design and SE architecture design | Phase 4: 2 weeks | Consistency checks between previous stage's artifacts. Screen designs and software architecture designs. |
| 10 (SE) | | | |
| 11 (UE) | Usability specs, prototyping, formative evaluation. Software detailed design, prototyping (**T & T**) | Phase 5: 2 weeks | Usability specs, prototypes, software detailed design |
| 12 (SE) | | | |
| 13 (Both) | Course wrap-up | Phase 6: 1 week | Evaluation results and analysis |
| 14 | Team presentations of project, lessons learned, problems found, constraints, dependencies, solutions to these issues. Feedback. **(Lessons learned)** | Phase 7: 2 weeks | Final report with issues and proposed solutions. |
| 15 | | | |

from the HCI and SE domains. For teaching oriented schools, there is a need for textbooks that can be used by CS professors to teach a course like this. In the former case, there is a requirement for substantial collaboration to teach integrated courses because the different domains in the same course should be taught by the professors from that particular domain.

- **Training through simulated activities:** Courses that teach development process models, be it SE or UE, require class activities that simulate a real project and let the student apply the skills and techniques learned in association with the process model. This is not an easy undertaking because it can be difficult to collect such real life activities for use in the classroom. In the case of UE, most companies (e.g. Apple) see their development models as one of their assets; something that gives them a competitive advantage, and often do not want to divulge how they develop their products.
- **Collaboration among students:** Students must learn how to collaborate with the other lifecycle subgroups. A good way to train students in this collaboration is to assign SE and UE roles (as described in the previous section) during the simulated projects and have them function in parallel, like in a real project. Each group should be asked to generate artifacts in collaboration with the opposite group.

### Textbooks

Even though some text books do address a few of the issues noted above [8], there is a dire need for textbooks that cover a substantial number of the topics discussed here. This lack of textbooks tailored for integrated curricula is a problem that cannot be solved overnight. Until the availability of such textbooks, the CS departments should prepare their own syllabi by picking the topics from 'pure' SE and UE textbooks that are available.

### SUMMARY

In this paper we provide an argument that software system failure rate can to some extent, be attributed to the way computer science and software engineering curricula are offered in universities. We present motivation and argue that there is a strong need for cross pollinating HCI and SE curricula to better equip the students for the realities of developing interactive software. We describe the main barriers to an integrated development process model and propose four major themes for a CS curriculum, namely, process models and components, techniques and tools, lessons learned, and applications. We then provide a detailed blueprint for a semester long course to teach the integrated system development methodology by using one possible combination of the four themes. We hope that by addressing this problem in the academia first, over time our students will have an impact in the industrial practice of developing interactive applications.